\documentclass[10pt,conference]{IEEEtran}
\IEEEoverridecommandlockouts
\usepackage{cite}
\usepackage{amsmath,amssymb,amsfonts}
\usepackage{algorithmic}
\usepackage{hyperref}
\usepackage{soul}
\usepackage{xcolor}
\usepackage{graphicx}
\usepackage{caption}
\usepackage{subcaption}

\newcommand{\email}[1]{\href{mailto:#1}{#1}}
\newcommand{\nsf}[1]{\href{https://www.nsf.gov/awardsearch/showAward?AWD_ID=#1}{#1}}

\begin{document}

\title{Evaluation of Noise and Crosstalk\\in Neutral Atom Quantum Computers%
\thanks{\par$^\dagger$ This work was supported in part by NSF grant \nsf{2332406}.}
}

\author{
    \IEEEauthorblockN{Pranet Sharma}
    \IEEEauthorblockA{%
        Yale University\\
        \email{pranet.sharma@yale.edu}
    }
    \and
    \IEEEauthorblockN{Yizhuo Tan}
    \IEEEauthorblockA{%
        Yale University\\
        \email{yizhuo.tan@yale.edu}
    }
    \and
    \IEEEauthorblockN{Konstantinos-Nikolaos Papadopoulos}
    \IEEEauthorblockA{%
        Northwestern University\\
        \email{kpapado@u.northwestern.edu}
    }
    \and
    \IEEEauthorblockN{Jakub Szefer}
    \IEEEauthorblockA{%
        Northwestern University\\
        \email{jakub.szefer@northwestern.edu}
    }
}

\maketitle

\begin{abstract}
This work explores and evaluates noise and crosstalk in neutral atom quantum computers. Neutral atom quantum computers are a promising platform for analog Hamiltonian simulations, which rely on a sequence of time-dependent Hamiltonians to model the dynamics of a larger system and are particularly useful for problems in optimization, physics, and molecular dynamics. However, the viability of running multiple simulations in a co-located or multi-tenant environment is limited by noise and crosstalk. This work conducts an analysis of how noise faced by simulations changes over time, and investigates the effects of spatial co-location on simulation fidelity. Findings of this work demonstrate that the close proximity of concurrent simulations can increase crosstalk between them. To mitigate this issue, a Moving Target Defense (MTD) strategy is proposed and evaluated. The results confirm that the MTD is a viable technique for enabling safe and reliable co-location of simulations on neutral atom quantum hardware.
\end{abstract}


\section{Introduction}

The field of quantum computing has transitioned from a realm of theoretical promise to one of tangible, rapidly advancing hardware. Among the diverse modalities being pursued, neutral atom quantum computers have distinguished themselves as a compelling platform. Within these computers, atoms of alkali and alkaline earth metals such as rubidium, strontium, and cesium serve as qubits. These atoms are first confined in a magneto-optical trap, and are then loaded into optical tweezers \cite{wurtz2023aquila}, which are used to move the atoms around for computation by exploiting the Rydberg blockade effect~\cite{surface2012horsman}. Due to the constant reconfiguration of qubits, the time cost of a quantum gate on current neutral atom quantum systems is high. However, they are well-suited for the computational paradigm of analog Hamiltonian simulations (AHS), one of the first applications of quantum computing with  demonstrated quantum advantage when modeling problems in optimization, physics, and molecular dynamics~\cite{practical2022daley}~\cite{cubitt2018universal}~\cite{trivedi2024quantum}. 

To maximize the computational efficiency and throughput of neutral atom systems running these simulations, they can use multi-tenancy, or co-location. Here, the large number of available qubits on a single quantum ``chip" are partitioned to run multiple, independent simulations in parallel. However, before such parallel execution can be reliably implemented, an engineering challenge must be overcome: the effects of system noise and inter-simulation crosstalk. Uncontrolled noise can affect the quantum states of a simulation, while crosstalk -- unwanted quantum mechanical interactions between supposedly independent qubits -- can create spurious correlations, compromising the results. The impacts of crosstalk and the ways that it can be potentially weaponized in cloud-based quantum computers have been studied in the recent past~\cite{nguyen2024quantum}.
%~\cite{chen2014qubit}~\cite{zhao2023mitigation}~\cite{harper2024crosstalk}.  
However, not much work has been done toward understanding possible noise and crosstalk in neutral atom quantum computers, nor developing runtime protection against crosstalk.

%Towards this end, in this work, we used the QuEra Aquila 256-qubit neutral atom quantum computer accessed via qBraid Lab to study noise and crosstalk. To establish a baseline understanding of the system's inherent noise characteristics, we performed a comprehensive temporal and spatial noise evaluation. This involved executing a standardized, single AHS program repeatedly at different physical locations across the quantum computer's qubit array over a one-month period. This longitudinal study allowed us to map the device's error landscape, revealing that noise, while largely random in nature, is not uniform.

%We further performed a noise and crosstalk evaluation using two simulations running in parallel on the Aquila quantum computer. This was done to evaluate co-location or multi-tenant scenarios to understand how crosstalk could be affecting different simulations when they are co-located. We found that distance affects crosstalk between simulations, and keeping a larger fixed distance between simulations could mitigate the crosstalk. However, such a solution could be inefficient. As a result, we developed and tested a moving target defense that actively moves the simulations before and after their execution. We found that the moving target defense can mitigate the impact of noise and crosstalk, and could be one way of safely using neutral atom quantum computers with multiple simulations running in parallel.

\begin{figure*}[t]
    \centering
    \begin{subfigure}[b]{0.24\textwidth}
        \centering
        \includegraphics[width=\linewidth]{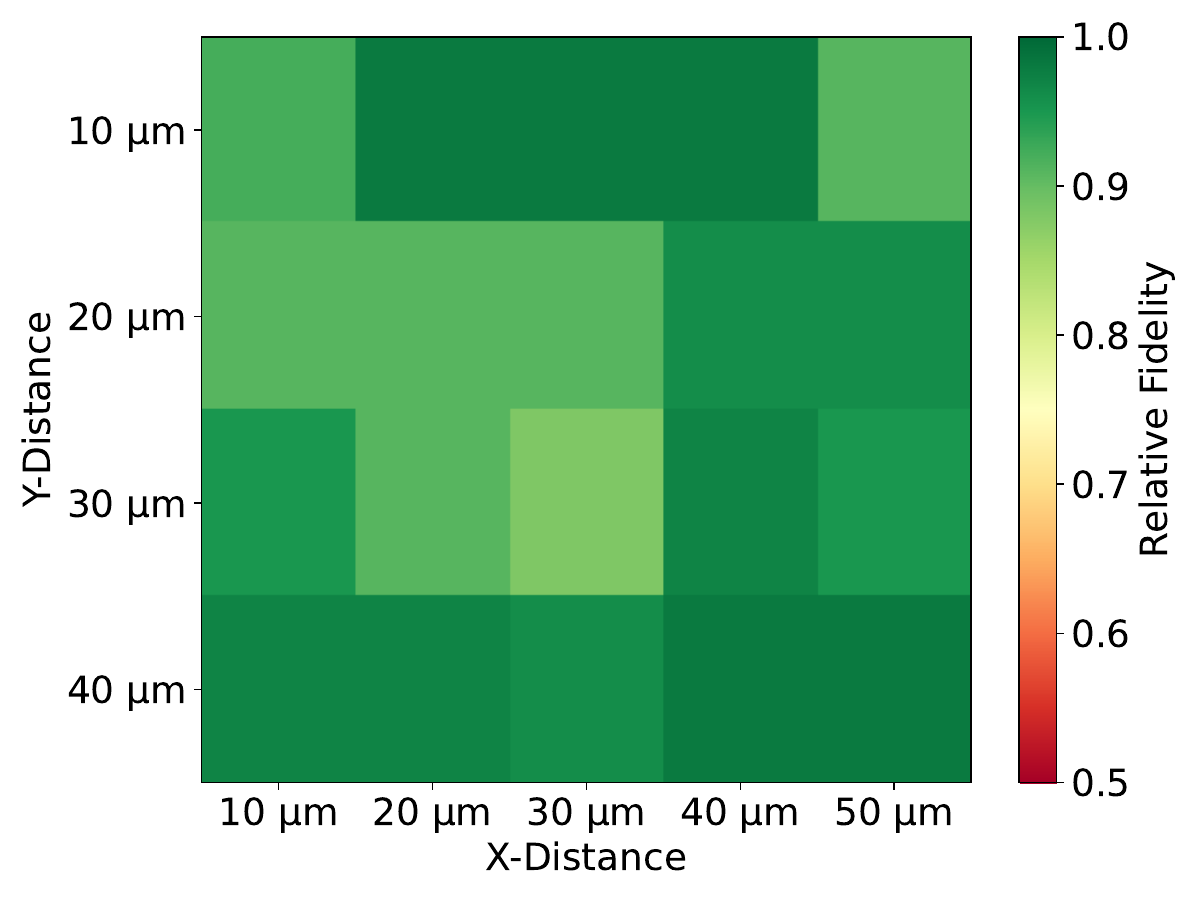}
        \caption{\small Test on May 3, 2025.}
        \label{fig_temporal_noise_heatmaps_subfig_1}
    \end{subfigure}
    %\hfill
    \begin{subfigure}[b]{0.24\textwidth}
        \centering
        \includegraphics[width=\linewidth]{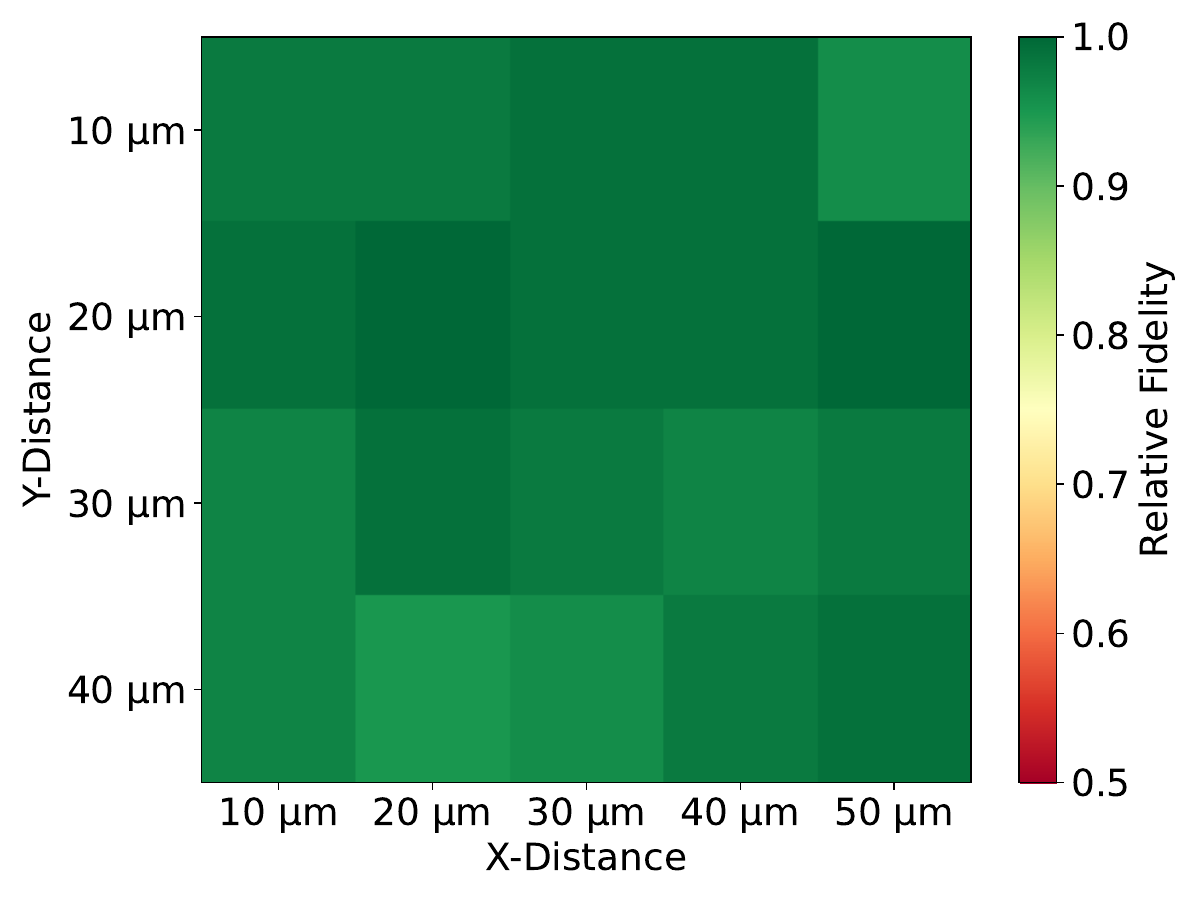}
        \caption{\small Test on May 10, 2025.}
        \label{fig_temporal_noise_heatmaps_subfig_2}
    \end{subfigure}
    %
    %\vskip\baselineskip
    %
    \begin{subfigure}[b]{0.24\textwidth}
        \centering
        \includegraphics[width=\linewidth]{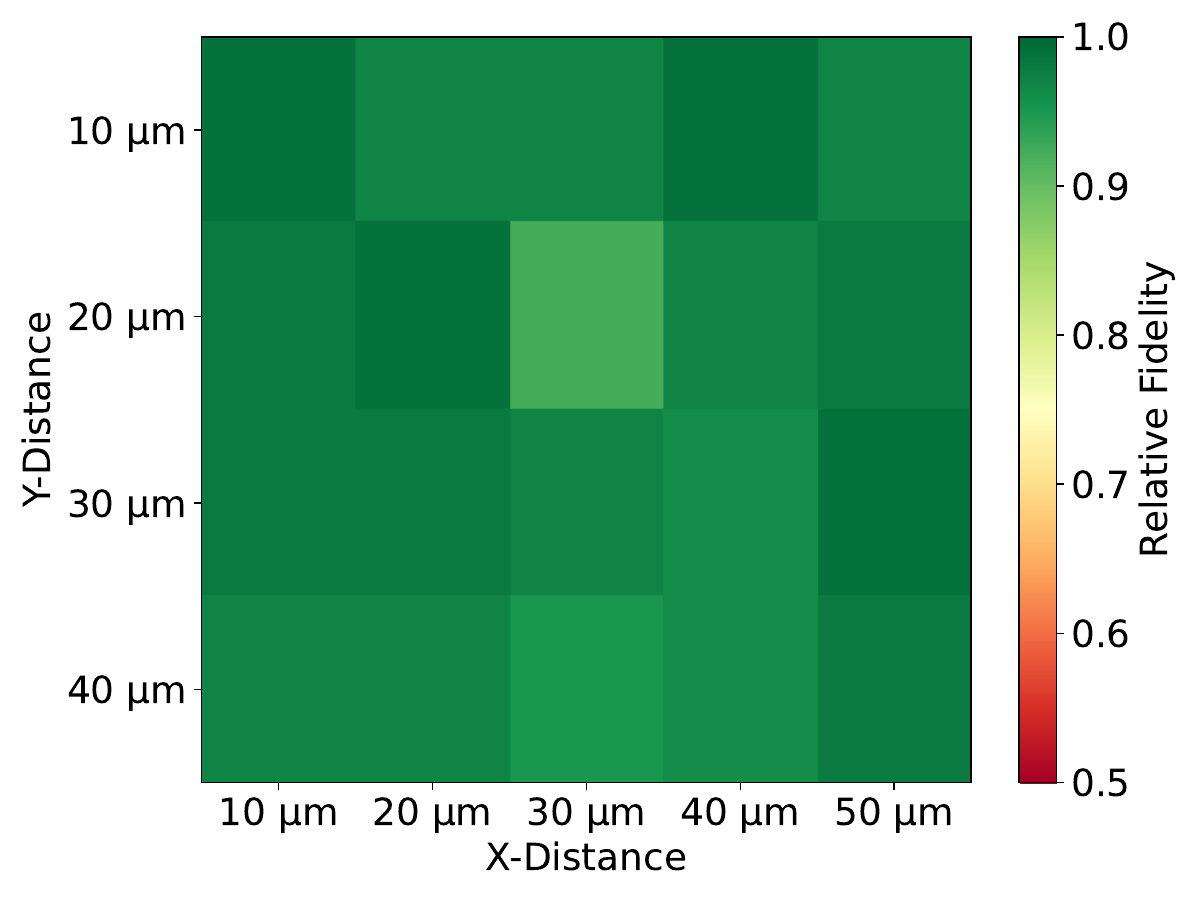}
        \caption{\small Test on May 17, 2025.}
        \label{fig_temporal_noise_heatmaps_subfig_3}
    \end{subfigure}
    %\hfill
    \begin{subfigure}[b]{0.24\textwidth}
        \centering
        \includegraphics[width=\linewidth]{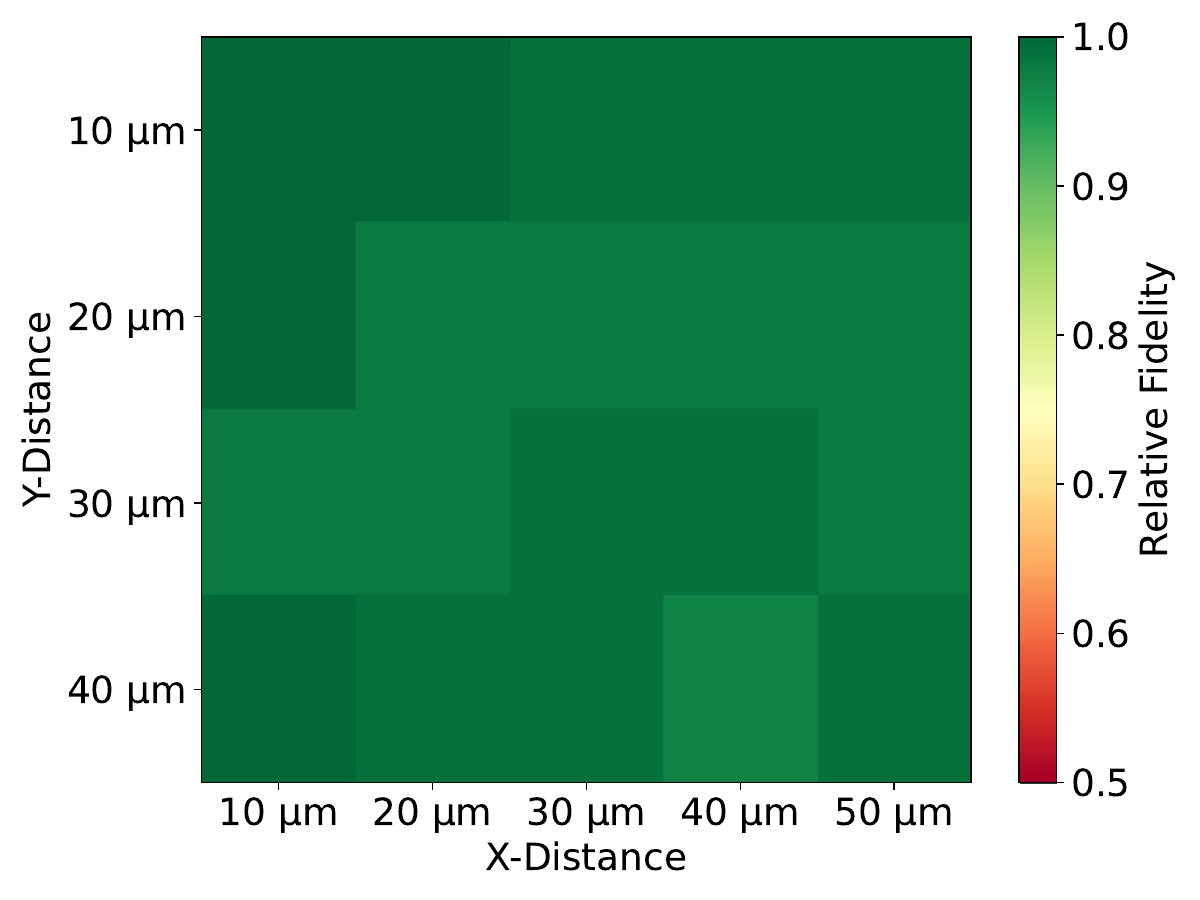}
        \caption{\small Test on May 24, 2025.}
        \label{fig_temporal_noise_heatmaps_subfig_4}
    \end{subfigure}
    
    \caption{\small Temporal noise evaluation on the Aquila quantum computer over period of four weeks in May, 2025. The darker green colors represent parts of the array that performed more closely to the determined control. Uniform coloring -- such as visible in Fig 1d -- implies less variance in final qubit counts across multiple runs, while non-uniform coloring -- such as visible in Fig 1a -- implies more variance.}
    \label{fig_temporal_noise_heatmaps}
\end{figure*}

\section{Simulation Framework and Control Setup}
\label{sec_simulation}

%Unlike for superconducting or trapped-ion machines, there has been little security analysis of neutral atom architectures. In our work we thus first characterize potential cross-talk and interference in a setting where the netural atom machine is used by different users, and then evaluate the defense.
%
% \subsection{Cross-talk Attack in Neutral Atom Hardware}
%

% \hl{In addition to the constructed attack described in this section, we have also conducted sample cross-talk attacks in past work [] and replicated those results here, demonstrating multi-hardware implementations of the moving target defense.}

%We chose the same register layout for $qs_A$ and $qs_V$. 
\subsection{Analog Hamiltonian Simulations}

 An AHS program relies on a sequence of time-dependent Hamiltonians governing the dynamics of the larger system. Neutral atom quantum computers execute AHS by mapping the time-dependent Hamiltonian evolution of their qubits (a quantum many-body Hamiltonian) to a real-life problem. Within this work, we used QuEra's neutral atom backends. QuEra offers three degrees of freedom to control the quantum many-body Hamiltonian:
\begin{equation}
\begin{split}
    H(t) &= \sum_{k=1}^N H_{\text{drive}, k}(t) \\
         &\quad + \sum_{k=1}^N H_{\text{shift}, k}(t) \\
         &\quad + \sum_{j=1}^{N-1}\sum_{k = j+1}^N H_{\text{vdW}, j, k}.
\end{split}
\end{equation}

\noindent where $H_{\text{drive}, k}(t)$ is the driving field, $H_{\text{shift}, k}(t)$ is the shifting field, and $H_{\text{vdW}, j, k}$ represents van der Waals interactions. The driving and shifting fields are \textit{externally} applied to the simulation, while the van der Waals field is \textit{internal} to the simulation and governs the interactions between each pair of qubits~\cite{wurtz2023aquila}. The driving field is uniformly applied to every qubit in the register. Therefore, in our experiments, we are more interested in the shifting field, as it can be applied directly to particular qubits as specified by the user. The impact of the shifting field creates an effect on the qubit known as ``detuning" -- shifting its frequency away from resonance -- which changes the van der Waals interactions of that qubit, thereby inducing a form of crosstalk~\cite{weiss2017quantum}.

The shifting field can be further defined as
\begin{equation}
    H_{\text{shift}, k}(t) = -\Delta_\text{local}(t)h_k \,n_k,
\end{equation}
where $\Delta_\text{local}(t)$ is the time-dependent magnitude of the frequency shift, and $h_k$ is the atom-dependent pattern~\cite{wurtz2023aquila}. A carefully chosen $\Delta_{\text{local}}$ value, therefore, can help simulate several real-world situations. On the other hand, a maliciously set shifting field can also be selected to induce the most disruption in nearby qubits -- which could be abused in the case of multiple simulations co-located on one quantum computer in a multi-tenant setting.

\subsection{Experimental Setup}

We used the same core simulation $qs_C$ in every experiment. We accessed two backends in this experiment -- Aquila, QuEra's cloud-accessible neutral atom QPU, and the Amazon Web Services (AWS) Braket AHS simulator for QuEra \cite{awsbraketservice}. Both backends were accessed via qBraid Lab \cite{qBraidLab_UserGuide_2025}. For our experiments, we set up a three-qubit register forming an equilateral triangle with side lengths of 5.5~{\textmu}m, selected as it is the starting atomic configuration provided by QuEra for AHS~\cite{wurtz2023aquila}. For our experiments delivered to the simulator, we applied a shifting field to the top qubit in this register defined by $\Delta_{\text{local}}=5\times10^7 \text{ rad/sec}$, which was 20 times the frequency of the uniform driving field. In our result, therefore, we expected the top qubit would not be excited due to its detuning, while the other two were likely to become entangled due to their van der Waals interactions. Within AHS, results are quantified by the atom count of each qubit after the conclusion of the simulation. Therefore, with the introduction of our shifting field, we can expect extremely low counts for the detuned qubit and high, even counts for the non-detuned qubits. In contrast, without the introduction of a shifting field, we expect more evenly distributed counts across all three qubits. We did not include this specified shifting field when submitting to Aquila, as it currently does not support local detuning without specialized Braket Direct access; in future work, we aim to replicate this setup through that service \cite{aws2025braket}.

We began by establishing our control: expected qubit counts for $qs_C$, determined by averaging individual qubit counts over multiple runs on the simulator. In each experiment, we defined \textit{relative fidelity} as the averaged percent difference in final qubit counts for each qubit in our experimentally run simulations from the control. The closer the relative fidelity is to 1, the better the backend performed when executing the~simulation.

\section{Temporal Noise Evaluation}

To perform the temporal noise evaluation, we submitted $qs_C$ for execution repeatedly, shifting the entire register by 10 {\textmu}m -- roughly double the default set by QuEra, therefore an interesting scale to explore -- to map a 40 {\textmu}m by 50 {\textmu}m area. We repeated this process every week for four weeks, aiming to understand how the noise changed over time. To improve our understanding of how simulations would vary, we additionally factored in the average qubit counts of all 20 executions when determining our expected counts to calculate relative fidelity. Figure \ref{fig_temporal_noise_heatmaps} contains a heatmap of relative fidelity for $qs_C$ executed within each part of the 40 {\textmu}m by 50 {\textmu}m area over the course of four weeks.

 We found that the noise remains mostly random. These observations are consistent with expectations regarding neutral atom platforms, where ambient temperature fluctuations, beam alignment shifts, and optical imperfections contribute to time-varying system noise \cite{wintersperger2023neutral}.

%This technique of periodic collection of noise data can provide a complementary diagnostic tool to standard benchmarking practices. Researchers and operators can map out noise variations within the device. As such, spatial noise monitoring can be considered an addition to the toolkit for monitoring long-term quantum computer performance and reliability.

\section{Spatial Noise Evaluation}

\begin{figure*}[t]
    \centering
    \includegraphics[width=0.95\linewidth]{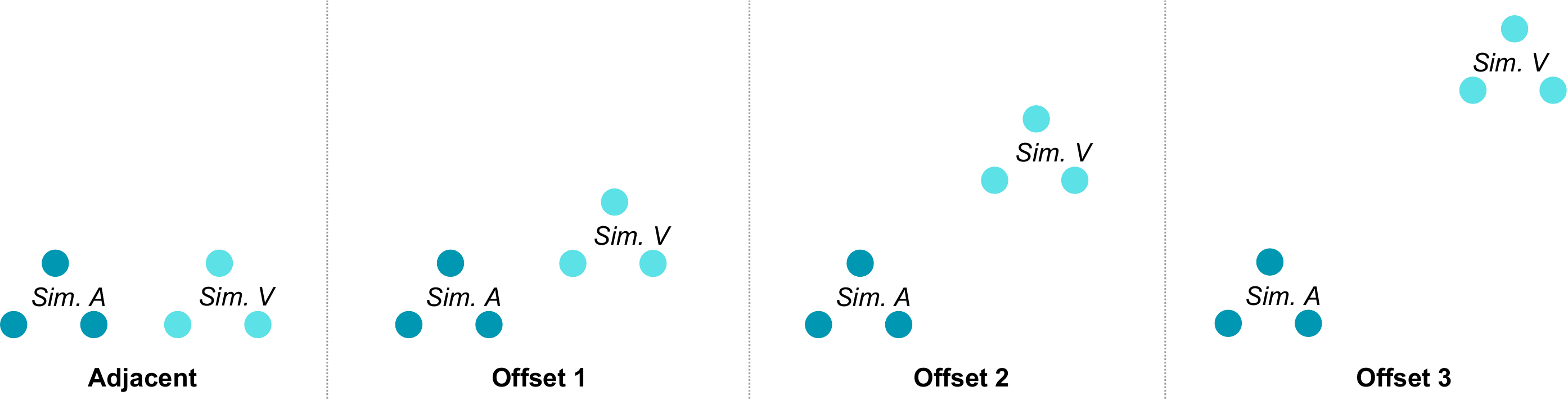}
    \caption{\small Spatial noise evaluation settings tested used this work. ``Adjacent'' configuration set the simulations 4~{\textmu}m apart. ``Offset 1'' set the qubits 5~{\textmu}m apart, ``Offset 2'' set the qubits 6~{\textmu}m apart, and ``Offset 3'' set the qubits 7~{\textmu}m apart. The figure is not to scale. The diagonal Euler distance between the qubits was used as the independent variable to control for horizontal or vertical biases in crosstalk.}
    \label{fig_co_location_distances}
\end{figure*}

To perform the spatial noise evaluation, we used two AHS with the same setup as $qs_C$: $qs_V$ and $qs_A$. We moved $qs_V$ an Euler distance of 1~{\textmu}m at a time from the co-located simulation $qs_A$.
%The ``Adjacent'' configuration set the simulations 4~{\textmu}m apart (the minimum allowed by QuEra). ``Offset 1'' set the qubits 5~{\textmu}m apart, ``Offset 2'' set the qubits 6~{\textmu}mapart, and ``Offset 3'' set the qubits 7~{\textmu}m apart. 
Figure~\ref{fig_co_location_distances} shows the different co-location settings.

%In the second phase, we simulated a cross-talk attack. The QuEra Aquila computer mandates a minimum atomic distance between qubits of 4 $\mu$m. This served as our minimum attack distance. We submitted the second simulation ($qs_A$) at the same time as the first one ($qs_V$), separating each qubit from its counterparts by $4$ $\mu$m. We then ran multiple trials and averaged the counts of the atoms after the conclusion of both simulations. 

%Currently, public cloud-based neutral atom quantum computers do not offer co-location. However, a growing body of academic work, e.g.,~\cite{das2019case}, is exploring co-location as a means of increasing the utility of quantum computers. Consequently, our spatial noise evaluation can help understand possible issues with fidelity and also how to mitigate the potential security threats of co-location.

Table~\ref{tab_co_location_fidelity} shows the relative fidelity of $qs_V$ for the different co-location distances tested. Our evaluation reveals a relationship between the physical separation of co-located simulations and the relative fidelity change observed. We found that the relative fidelity of $qs_V$ at offsets $5$ \textmu m or more behaves roughly linearly with the distance between it and $qs_A$. Further, in our experiment, the greatest fidelity disruption occurred when $qs_A$ was positioned 5~{\textmu}m away from $qs_V$. At smaller separations, the measurements were very variable. As $qs_V$ continued to move further away from $qs_A$, the relative fidelity of $qs_V$ approached 1, indicating that distances of greater than $8$~{\textmu}m between simulations appear to be more resilient to the effects of crosstalk. This makes $8$~{\textmu}m a promising lower bound to separate simulations if co-location is to be safely implemented.

%\begin{figure}[t]
%    \centering
%    \includegraphics[width=0.75\linewidth]{example-image-a}
%    \caption{\small Simulation fidelity at different spatial separation distances.}
%    \label{fig_co_location_fidelity}
%\end{figure}

\begin{table}[t]
\centering
\caption{\small Simulation fidelity at different spatial separation distances. Fidelity is relative to having no co-location.}
\label{tab_co_location_fidelity}
\begin{tabular}{lc}
\hline
\textbf{Offset} & \textbf{Relative Fidelity}\\
\hline
Adjacent (4~{\textmu}m)  & $0.963 \pm 0.026$ \\
Offset 1 (5~{\textmu}m)   & $0.882 \pm 0.010$ \\
Offset 2 (6~{\textmu}m)  & $0.925 \pm 0.033$ \\
Offset 3 (7~{\textmu}m)  & $0.984 \pm 0.009$ \\
\hline
\end{tabular}
\end{table}

\section{Moving Target Defense}

\begin{figure*}[t]
    \centering
    \includegraphics[width=0.70\linewidth]{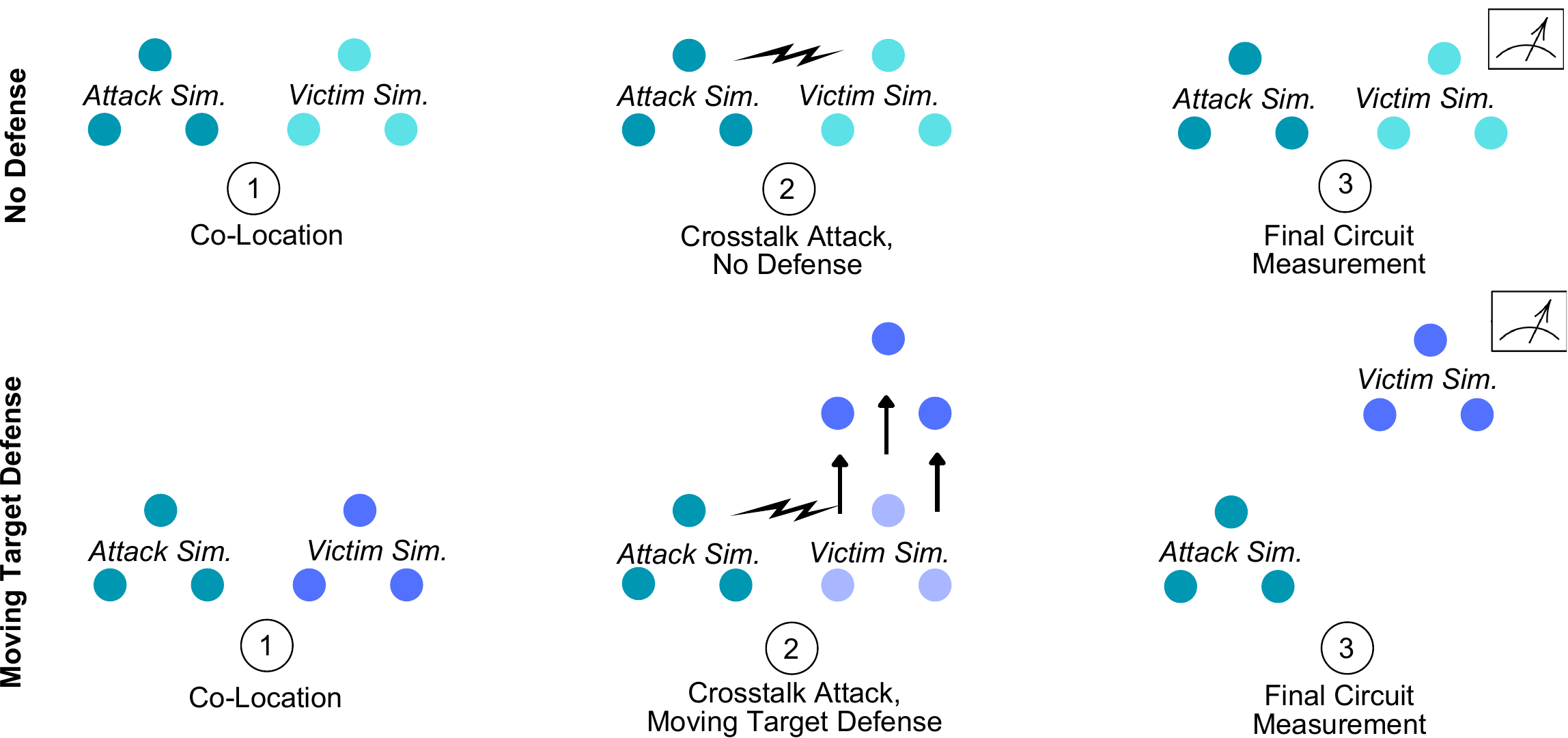}
    \caption{\small Moving target defense overview. The figure is not to scale. At each step, the victim circuit was moved $4$ \textmu m. The simulation can optionally be moved further after measurement.}
    \label{fig_moving_target}
\end{figure*}

Having a fixed and large separation between simulations reduces crosstalk, but may create inefficiencies in how the quantum computer is utilized. At the same time, having a fixed and short separation provides for better utilization, but simulations become prone to crosstalk. As a possible solution to this problem, we propose a Moving Target Defense (MTD) for neutral atom quantum computers, inspired by classical work~\cite{cho2020toward}. The idea for moving target defense on neutral atom architectures is to physically move the simulation within the register before and optionally after execution steps of the simulation. This idea is shown in Figure~\ref{fig_moving_target}.

To validate that our proposed moving target defense can effectively mitigate the risks of a crosstalk attack, we first demonstrate a crosstalk attack on the AHS simulator\footnote{When we are able to access local detuning within Aquila, either through Braket Direct or an expansion of features generally available, we aim to replicate this experiment on a QPU.}.  We accomplish this by simultaneously submitting two simulations within the same register, with $qs_A$ serving as the ``attack simulation" and $qs_V$ serving as the ``victim simulation". As the location and encoding of the qubits are both fully programmable on neutral atom quantum hardware, and it is easy to control the positions of two simulations at once, thereby ensuring our control over the procedure from the attack side and the defense side. Due to the utilization of neutral atom hardware in several modern-day simulation use cases with sensitive data involved, demonstrating a working defense has significance for the current era of quantum computing~\cite{wintersperger2023neutral}.

\begin{figure}[t]
    \centering
\includegraphics[width=0.55\linewidth]{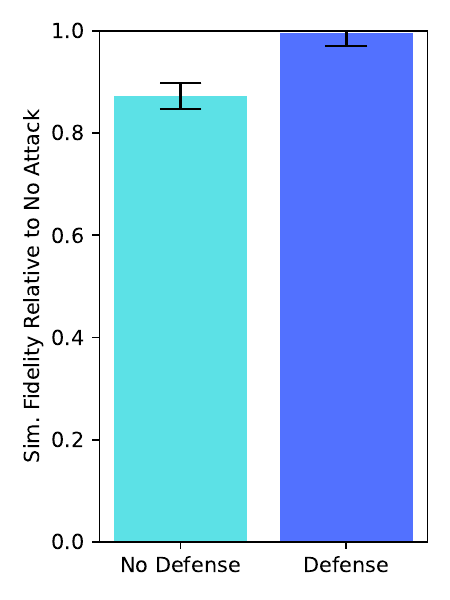}
    \caption{\small Fidelity without and with moving target defense. Initially, the separation between victim and attacker was $5$ \textmu m.}
    \label{fig_moving_target_fidelity}
\end{figure}

Within our moving target defense experiment, we ran $qs_A$ and $qs_V$ with a separation of 5 \textmu m for ``No Defense" data, and moved $qs_V$ away from $qs_A$ twice in 4 \textmu m intervals for the ``Defense" data. Figure~\ref{fig_moving_target_fidelity} shows that the fidelity of $qs_V$ with co-location decreases without the implementation of a moving target defense. Further, with the implementation of a moving target defense, $qs_V$ had a relative fidelity of $0.995\pm0.02$, indicating close to expected performance, well within the bounds of noise-induced fidelity changes. This demonstrates that a moving target defense can mitigate the impact of noise and crosstalk, and could be one way to safely operate neutral atom quantum computers with multiple simulations running in parallel.

\section{Related Work}

While there have been works that review the reliability of neutral atom quantum computers~\cite{wintersperger2023neutral}~\cite{wagner2024benchmarking}, this work is to the best of our knowledge the first to study reliability in a cloud setting and examine potential attacks targeting cloud-based neutral atom quantum computers. In the limited space, we present security-minded related work on quantum computers.

%\subsection{Crosstalk attacks}

In the context of multi-tenant quantum computers based on superconducting qubits, many attacks utilize crosstalk to leak information. Recent work introduces QubitHammer~\cite{2504.07875}, a set of attacks that use custom pulses to degrade the fidelity of victim circuits—even when the attacker and victim are placed far apart on the quantum topology. The authors also show that current mitigation techniques such as dynamical decoupling, disabling specific qubits (e.g., qubit $0$), crosstalk-aware qubit allocation, and active padding are largely ineffective. A similar row hammer-style attack is demonstrated in~\cite{2503.21650}, where repeated two-qubit gate operations are used to flip the state of a neighboring victim qubit. These and related works focus usually on superconducting quantum computer crosstalk, while our work explores neutral atom architectures.

%Another paper~\cite{choudhury2025crosstalk} exploits crosstalk side-channel leakage to extract information about a victim's quantum circuit in a shared NISQ system. The adversary runs a crosstalk estimation circuit co-located with the victim’s, detecting the number and placement of CNOT gates, which are highly crosstalk-prone. They train a GCN-based model using features such as qubit count, CNOT density per qubit, gate timing, and pairwise CNOT patterns to accurately reconstruct the victim’s circuit.

%Another study~\cite{ash2020analysis} analyzes crosstalk error rates and demonstrates a fault-injection attack that degrades output fidelity and can result in denial-of-service conditions.
%Also, work in~\cite{saki2021qubit} leverages the state-dependent nature of measurement and readout error probabilities as a side channel in multi-programming setups, allowing adversaries to infer the output of victim circuits with high accuracy.

%\subsection{Other types of attacks}

Other works target the quantum cloud infrastructure more broadly. For example, fingerprinting attacks, described in ~\cite{mi2021short}, leverage crosstalk-induced error patterns to uniquely identify specific quantum devices. 
%Some studies explore timing side channels instead of crosstalk. In~\cite{lu2024quantum}, attacker circuits are inserted between executions of the victim’s circuit to estimate timing characteristics, successfully revealing the circuit, secret oracle, and quantum device identity—though not the ansatz or qubit mapping.
%In~\cite{roy2024hardware}, small hardware Trojan circuits are inserted during compilation to intentionally degrade output fidelity.
Finally, researchers have demonstrated that power side-channel attacks can be used to recover the sequence of quantum gates being executed. In~\cite{xu2023exploration}, power traces from the control electronics are analyzed to infer qubit control pulses and reverse-engineer the underlying algorithm. We have not explored crosstalk or side channels in control electronics in neutral atom architectures, and this could be a valuable future~direction.

\section{Conclusion}

This work investigated noise and crosstalk in neutral atom quantum computers, specifically in the context of cloud co-tenancy. Experimental results demonstrated that spatial co-location resulted in crosstalk-induced noise. In response, a moving target defense mechanism was proposed and evaluated on neutral atom quantum computing platforms. With the high relative fidelity of circuits protected by the moving target defense, this technique may be one strategy for enabling secure co-location and co-tenancy of simulations on neutral atom quantum computers.

\bibliographystyle{IEEEtran}
\bibliography{refs/refs}

\begin{thebibliography}{10}
\providecommand{\url}[1]{#1}
\csname url@samestyle\endcsname
\providecommand{\newblock}{\relax}
\providecommand{\bibinfo}[2]{#2}
\providecommand{\BIBentrySTDinterwordspacing}{\spaceskip=0pt\relax}
\providecommand{\BIBentryALTinterwordstretchfactor}{4}
\providecommand{\BIBentryALTinterwordspacing}{\spaceskip=\fontdimen2\font plus
\BIBentryALTinterwordstretchfactor\fontdimen3\font minus \fontdimen4\font\relax}
\providecommand{\BIBforeignlanguage}[2]{{%
\expandafter\ifx\csname l@#1\endcsname\relax
\typeout{** WARNING: IEEEtran.bst: No hyphenation pattern has been}%
\typeout{** loaded for the language `#1'. Using the pattern for}%
\typeout{** the default language instead.}%
\else
\language=\csname l@#1\endcsname
\fi
#2}}
\providecommand{\BIBdecl}{\relax}
\BIBdecl

\bibitem{wurtz2023aquila}
J.~Wurtz, A.~Bylinskii, B.~Braverman, J.~Amato-Grill, S.~H. Cantu, F.~Huber, A.~Lukin, F.~Liu, P.~Weinberg, J.~Long \emph{et~al.}, ``Aquila: Quera's 256-qubit neutral-atom quantum computer,'' \emph{arXiv preprint arXiv:2306.11727}, 2023.

\bibitem{surface2012horsman}
D.~Horsman, A.~G. Fowler, S.~Devitt, and R.~Van~Meter, ``Surface code quantum computing by lattice surgery,'' \emph{New Journal of Physics}, vol.~14, no.~12, p. 123011, 2012.

\bibitem{practical2022daley}
A.~J. Daley, I.~Bloch, C.~Kokail, S.~Flannigan, N.~Pearson, M.~Troyer, and P.~Zoller, ``Practical quantum advantage in quantum simulation,'' \emph{Nature}, vol. 607, no. 7920, pp. 667--676, 2022.

\bibitem{cubitt2018universal}
T.~S. Cubitt, A.~Montanaro, and S.~Piddock, ``Universal quantum hamiltonians,'' \emph{Proceedings of the National Academy of Sciences}, vol. 115, no.~38, pp. 9497--9502, 2018.

\bibitem{trivedi2024quantum}
R.~Trivedi, A.~Franco~Rubio, and J.~I. Cirac, ``Quantum advantage and stability to errors in analogue quantum simulators,'' \emph{Nature Communications}, vol.~15, no.~1, p. 6507, 2024.

\bibitem{nguyen2024quantum}
H.~T. Nguyen, P.~Krishnan, D.~Krishnaswamy, M.~Usman, and R.~Buyya, ``Quantum cloud computing: A review, open problems, and future directions,'' \emph{arXiv preprint arXiv:2404.11420}, 2024.

\bibitem{weiss2017quantum}
D.~S. Weiss and M.~Saffman, ``Quantum computing with neutral atoms,'' \emph{Physics Today}, vol.~70, no.~7, pp. 44--50, 2017.

\bibitem{awsbraketservice}
{Amazon Web Services}, ``{Amazon Braket: Quantum computing on AWS},'' \url{https://aws.amazon.com/braket/}, 2025.

\bibitem{qBraidLab_UserGuide_2025}
{qBraid}, ``qbraid lab user guide,'' \url{https://lab.qbraid.com/}, 2025.

\bibitem{aws2025braket}
{Amazon Web Services}, ``Amazon braket features (with braket direct),'' \url{https://aws.amazon.com/braket/features/#Braket_Direct}, Jul 2025.

\bibitem{wintersperger2023neutral}
K.~Wintersperger, F.~Dommert, T.~Ehmer, A.~Hoursanov, J.~Klepsch, W.~Mauerer, G.~Reuber, T.~Strohm, M.~Yin, and S.~Luber, ``Neutral atom quantum computing hardware: performance and end-user perspective,'' \emph{EPJ Quantum Technology}, vol.~10, no.~1, p.~32, 2023.

\bibitem{cho2020toward}
J.-H. Cho, D.~P. Sharma, H.~Alavizadeh, S.~Yoon, N.~Ben-Asher, T.~J. Moore, D.~S. Kim, H.~Lim, and F.~F. Nelson, ``Toward proactive, adaptive defense: A survey on moving target defense,'' \emph{IEEE Communications Surveys \& Tutorials}, vol.~22, no.~1, pp. 709--745, 2020.

\bibitem{wagner2024benchmarking}
N.~Wagner, C.~Poole, T.~Graham, and M.~Saffman, ``Benchmarking a neutral-atom quantum computer,'' \emph{International Journal of Quantum Information}, vol.~22, no.~04, p. 2450001, 2024.

\bibitem{2504.07875}
\BIBentryALTinterwordspacing
Y.~Tan, N.~Choudhury, K.~Basu, and J.~Szefer, ``Qubithammer attacks: Qubit flipping attacks in multi-tenant superconducting quantum computers,'' 2025. [Online]. Available: \url{https://arxiv.org/abs/2504.07875}
\BIBentrySTDinterwordspacing

\bibitem{2503.21650}
\BIBentryALTinterwordspacing
F.~Almaguer-Angeles, P.~R. Dieguez, A.~S. H., and M.~Pawłowski, ``Hacking quantum computers with row hammer attack,'' 2025. [Online]. Available: \url{https://arxiv.org/abs/2503.21650}
\BIBentrySTDinterwordspacing

\bibitem{mi2021short}
A.~Mi, S.~Deng, and J.~Szefer, ``Short paper: Device- and locality-specific fingerprinting of shared nisq quantum computers,'' in \emph{Hardware and Architectural Support for Security and Privacy}, 2021.

\bibitem{xu2023exploration}
C.~Xu, F.~Erata, and J.~Szefer, ``Exploration of power side-channel vulnerabilities in quantum computer controllers,'' in \emph{Conference on Computer and Communications Security}, 2023.

\end{thebibliography}

\end{document}